\documentclass[9pt,conference]{IEEEtran}
\usepackage{amssymb,amsthm,amsmath,array}
\usepackage{graphicx}
\usepackage[caption=false,font=footnotesize]{subfig}
\usepackage{xspace}
\usepackage[sort&compress, numbers]{natbib}
\usepackage{stmaryrd}
\usepackage{units}
\usepackage{xcolor}
\usepackage{mathtools}
\usepackage{float}
\usepackage{textcomp}
\DeclarePairedDelimiter{\norm}{\lVert}{\rVert} 
\DeclareMathOperator*{\argmin}{arg\,min}

\begin{document}
\title{Practical Low-density Codes for PB-ToF Sensing}
\author{\IEEEauthorblockN{
        Alvaro Lopez Paredes, and 
        Miguel Heredia Conde
    }
    \IEEEauthorblockA{
        Center for Sensor Systems (ZESS), University of Siegen, Germany\\
        alvaro.lparedes@uni-siegen.de, heredia@zess.uni-siegen.de}
}
\maketitle

\begin{abstract}
An indirect Pulse-based Time-of-Flight camera can be modelled as a linear sensing system in which the target's depth is recovered from few measurements through a sensing matrix formed by a set of demodulation functions. Each demodulation function is the result of the convolution of a (0,1)-binary code and a cross-correlation function which models the entire modulation-demodulation process. In this paper, we present a practical scheme for the construction of the sensing matrix which relies on the optimization of the coherence, and is based on low-density codes. We demonstrate that our methodology eliminates the intrinsic variability of random and pseudo-random approaches, and allows for the recovery of the target's depth in a grid much finer than the number of distinct elements in the binary codes.


\thanks{This project has received funding from the European Union’s Horizon 2020 research and innovation programme under the Marie Skłodowska-Curie grant agreement No 860370-Menelaos\_NT project.}
\end{abstract}


\section{Introduction}

\begin{figure}[!b]
\centering
 \vspace{-15pt}
 \includegraphics[width=\linewidth]{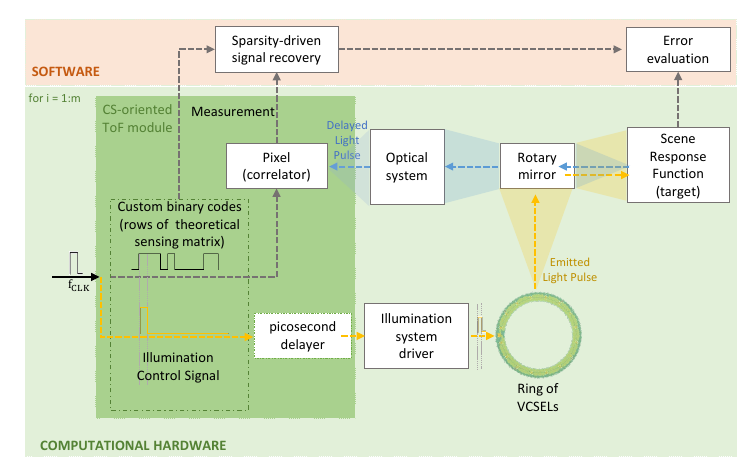}
 \vspace{-20pt}
\caption{Coded demodulation scheme for the PB-ToF camera designed at ZESS destined to very-wide-area applications.} 
\label{fig01}
\end{figure}

Time-of-Flight (ToF) systems generate a 3D representation of the surrounding space by calculating the time of the return-trip of a light signal from the sensor to the target. In indirect Pulse-based (PB)-ToF systems, as the one presented in Fig.~\ref{fig01} for our prototype, the target's depth is recovered from  samples of the correlation of the scene response function $[x^j]_{j=1}^n$ and a number of demodulation functions generated in the camera $[[a_i^j]_{i=1}^m]_{j=1}^{n}$, such that $y_i=\sum_{j=1}^{n}a_{i}^{j}\cdot x^{j}$ with $i=1:m$, and $m\ll n$ \cite{Bhandari1}. Retrieving $\vec{x}$ from $\vec{y}$ is an under-determined problem, and translates into a linearly-constrained $\ell_0$-minimization problem \eqref{eq01} considering the sparsity (or compressibility) of the scene response function.

\begin{equation}
  \hat{\vec{x}}= \underset{\vec{x}} \argmin  {\norm{\vec x}_0}, \quad \text{s.t.:} \quad \vec {y}= \pmb{A} \cdot \vec{x}
\label{eq01}
\end{equation}

The demodulation function $\left[\vec{a}_i\right]_{i=1}^m$ is the result of the convolution of a (0,1)-binary code and the cross-correlation function, that models the entire modulation-demodulation process and can be approximated by a Gaussian filter with $\sigma\sim\mathcal{O}\left(\unit[]{ns}\right)$. The discrete modelling of $\left[\vec{a}_i\right]_{i=1}^m$ in a grid much finer than the one established by the (0,1)-binary codes, i.\,e., $n_\mathrm{samples}\gg n$, allows for a much more precise localization of the target, i.\,e., to reach time (depth) super-resolution ($\gamma_\mathrm{SR}=\frac{n_\mathrm{samples}}{n}>1$). In addition, the reduction of the number of measurements $m$ becomes a key parameter to approach real-time, as the pixels require a certain amount of time $t_\mathrm{exp}\sim\mathcal{O}\left(\unit[]{ms}\right)$ to gather enough light. The construction of $\pmb{A}$ under these premises is not an easy task, as oversampling translates into high coherence \eqref{eq2} ($\mu=1$) and, therefore, to an ill-posed problem.

\begin{equation}
    \mu=\max{\left(\frac{\lvert{\vec{a}^{k^\intercal}} \cdot \vec{a}^{j}\rvert}{\norm{\vec{a}^k}_2\norm{\vec{a}^j}_2}\right)} \quad \mathrm{with} \quad k\neq j
    \label{eq2}
\end{equation}

\begin{figure}[htbp]
\centering
 \includegraphics[width=\linewidth]{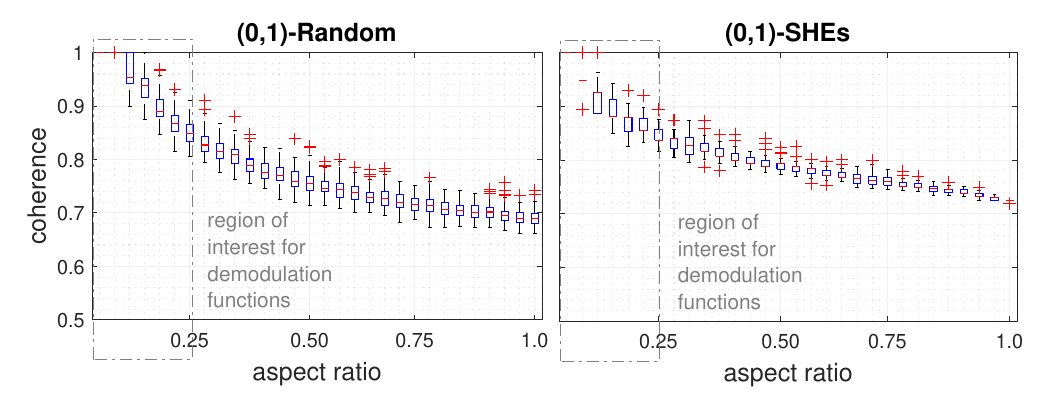}
 \vspace{-15pt}
\caption{Coherence vs aspect ratio of sensing matrices generated via random (0,1)-binary codes (left) and Scrambled Hadamard Ensembles (SHEs) with (0,1)-rescaling (right) over $n_\mathrm{real}=100$ realizations.} 
\label{fig02}
\end{figure}

Random and Hadamard-derived (0,1)-binary codes may be used in the modulation/demodulation schemes \cite{lopezparedes_2}, as the resulting sensing matrices ensure good recoverability properties for moderate aspect ratios ($\eta=\frac{m}{n}$), as shown in Fig.~\ref{fig02}. However, these schemes may lead to $\mu=1$ for $\eta\ll1$. To solve that issue, some deterministic approaches such as the ones which make use of Low-Density Parity-Check (LDPC) codes via Progressive-Edge Growth (PEG) \cite{PEGlike} were explored in \cite{lopezparedes_2}. PEG works iteratively per columns, adding a fixed number of non-zero elements at the locations which locally maximize the girth of the associated Tanner graph. However, a sequence of locally optimal decisions may not lead to a global minimum and $\mu<1$ cannot always be guaranteed. Recently, Gupta \cite{Gupta_2018} and Barragan \cite{Gutierrez-Barragan19} proposed a family of novel functions built upon Hamiltonian cycles on hypercube graphs, similarly to Gray codes. In this work, we propose a practical methodology which allows for $\gamma_\mathrm{SR}>1$, and ensures good recoverability properties for $\eta\ll1$. In contrast to \cite{Gupta_2018,Gutierrez-Barragan19}, our methodology is built upon low-density codes, as they also minimize the integration of background light, and allow for a faster computation if a sparsity-oriented programming is adopted. 


\section{Methodology}

Our coding scheme, called gradient-Combinatorial (GComb) \cite{lopezparedes_2}, iteratively adds $[\vec{a}^j]_{j=1}^{n}$, and consists of the following steps:

\begin{enumerate}
    \item We select the location of the $n_\mathrm{deg}$ non-zero elements in each of the $n$ columns from their possible combinations without repetition in the $m$ rows. This ensures $\mu<1$ in the (0,1)-binary codes that the demodulation functions are built upon.
    
    \item In real operation, the non-zero transition time of binary control signals yields coincidences of rising and falling edges. This may lead to identical columns far apart from each other in $\pmb{A}$. We include an evaluation step when adding a new element to $\pmb{A}$. Firstly, we generate a new matrix $\pmb{A}_\mathrm{dif}$, given by the differences between the elements of adjacent columns, and evaluate its coherence ($\mu_\mathrm{dif}$). If the added non-zero element yields $\mu_\mathrm{dif}=1$, we remove it from the set of candidates. This evaluation step translates into a deterministic arrangement of $[\vec{a}^j]_{j=1}^{n}$, and can be applied to any other methodology such as LDPC-PEG \cite{lopezparedes_2}.
\end{enumerate}

\begin{figure}[htbp]
 \vspace{-5pt}
 \centering
 \includegraphics[width=\linewidth]{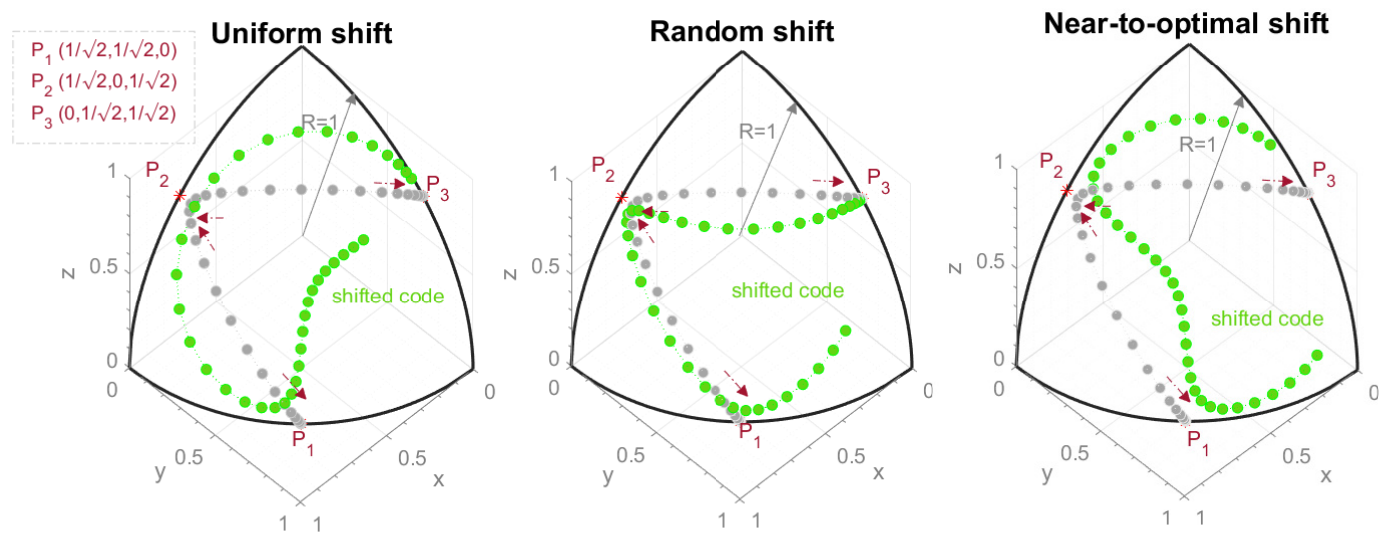}
 \vspace{-15pt}
\caption{Representation of normalized columns of the sensing matrices on the unit hyper-sphere on $\mathbb{R}^3$ for $n_\mathrm{steps}=14$, and $n_\mathrm{deg}=2$ when implementing uniform (left), random (center), and near-to-optimal (right) on-grid shifts \cite{LopezParedes_3}.} 
\label{fig03}
\end{figure}

GComb yields an upper bound for $\gamma_\mathrm{SR}$ such that $\mu<1$. We push this limit further by introducing near-to-optimal on-grid offsets (implemented through time delays) between $[\vec{a}_i]_{i=1}^{m}$. In our methodology \cite{LopezParedes_3}, given $\pmb{A}$, we sequentially work per rows, evaluate the impact of any possible on-grid offset, and select the one which maximizes the minimum inter-column distance. Our technique allows for a reduction of the number of highly-correlated columns, and outperforms other methodologies such as the ones based on uniform or random on-grid offsets. Fig.~\ref{fig03} presents a comparison of the distributions of $n_\mathrm{samples}=42$ samples  (columns of $\pmb{A}$) in the $\mathbb{R}^3$ hyper-sphere \cite{LopezParedes_3}. We observe an increase of the chordal distance between adjacent samples when the shifts are implemented, which yields a reduction of $\mu$.

\section{Numerical Simulations}
At ZESS, we have designed a rotary PB-ToF camera that allows us to implement the presented computational sensing approaches to bring long-range wide-area ToF imaging to a new level of accuracy \cite{lopezparedes_2}. The design parameters of our prototype used in this numerical simulation are shown in Table~\ref{tab01}.

\vspace{-5pt}
 \begin{table}[htbp]
    \centering
    \caption{Design parameters of our ToF camera}
    \label{table}
    \setlength{\tabcolsep}{4pt}
    \begin{tabular}
     {|p{140pt}|p{44pt}|p{42pt}|}
    \hline
    \textbf{Operational parameter} & \textbf{Acronym} & \textbf{Value}  \\  
    \hline  
    Modulation frequency (element bandwidth) & $f_\mathrm{m}$ &$\unit[448]{MHz}$  \\  
    \hline      
    Repetition frequency & $f_\mathrm{r}$ & $\unit[3.5]{MHz}$  \\  
    \hline 
    Number of sub-steps for super-resolution & $n_\mathrm{steps}$ & 8  \\  
    \hline
    Cross-correlation function & $\mathrm{FWHM}$ & $\unit[0.4]{ns}$  \\  
    \hline 
    Number of measurements & $m$ & 14  \\  
    \hline
     Number of non-zero elements per column & $n_\mathrm{deg}$ & 3  \\  
    \hline 
    \end{tabular}
    \label{tab01}
    \end{table} 

\begin{figure}[htbp]
\centering
 \includegraphics[width=\linewidth]{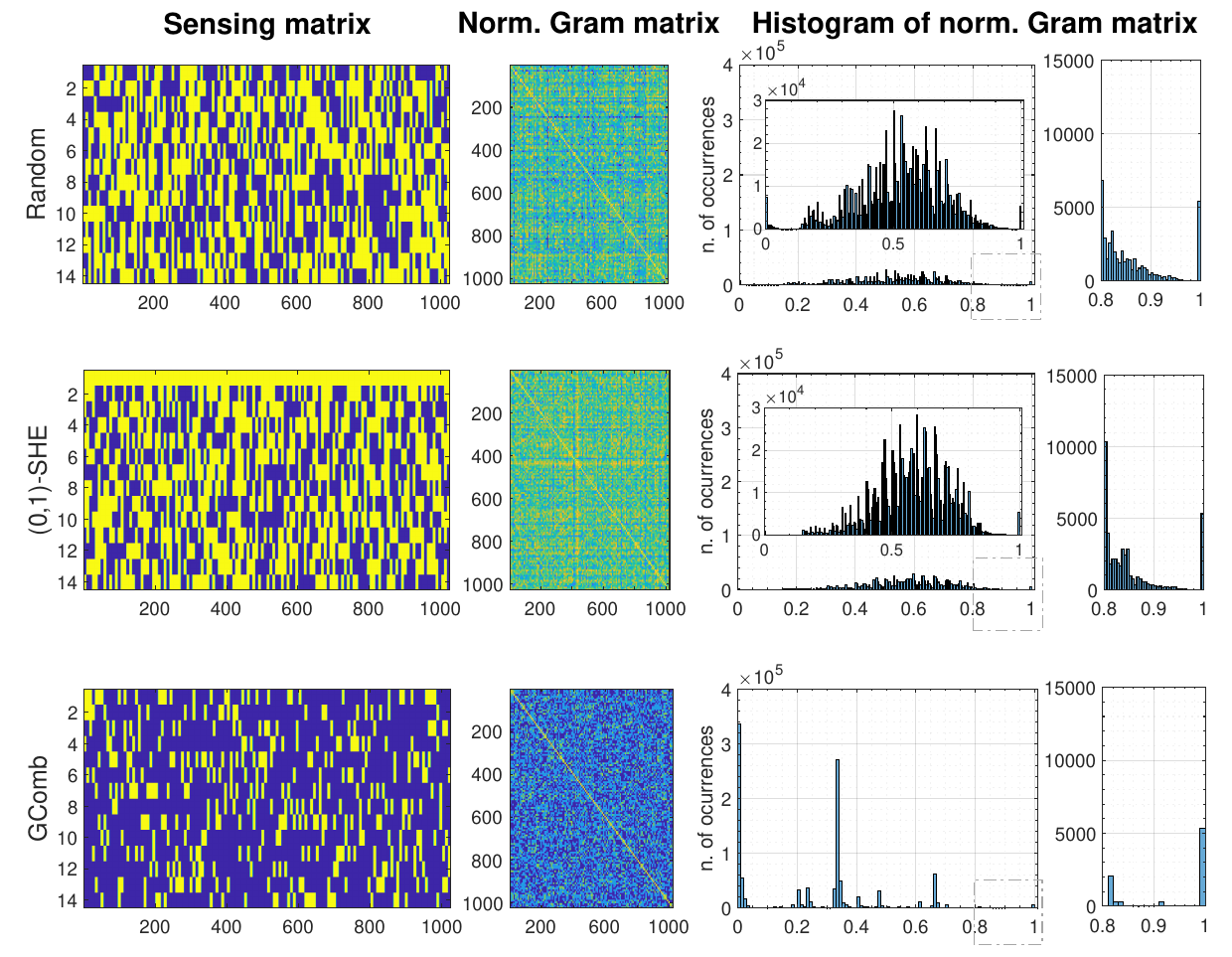}
\centering
 \vspace{-0pt}
 \includegraphics[width=\linewidth]{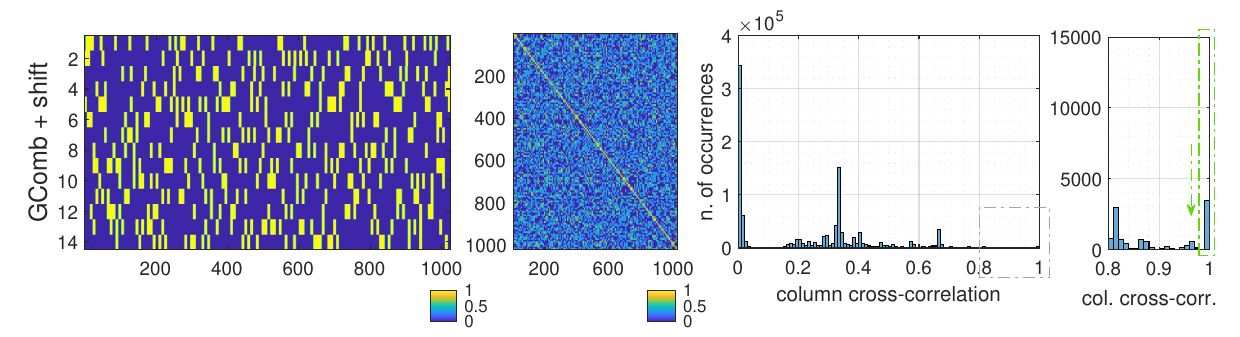}
 \vspace{-15pt}
\caption{Sensing matrices, normalized Gram matrices and histograms of column cross-correlations generated via random (0,1)-binary codes, Scrambled Hadamard Ensembles with (0,1)-rescaling, gradient-Combinatorial, and gradient-Combinatorial + near-to-optimal shifts. } 
\label{fig04}
\vspace{-5pt}
\end{figure}

Fig.\ref{fig04} shows the sensing matrices (left), the corresponding normalized Gram matrices (center), and the histograms of column cross-correlations (right) using random (0,1)-binary codes, (0,1)-Scrambled Hadamard Ensembles (SHEs), GComb, and GComb followed by the implementation of near-to-optimal on-grid shifts. We observe a drastic reduction of the number of highly-correlated columns, and a concentration of them in a narrow region close to the diagonal of the normalized Gram matrix for GComb. In addition, we identify a further reduction of the number of highly-correlated columns when introducing near-to-optimal shifts in a matrix built via GComb.

\section{Conclusion}
In this paper, we have presented a practical methodology for the generation of sensing matrices that makes use of low-density codes and ensures good coherence properties for very low aspect ratios. This allows for the use of these codes in close to real-time PB-ToF systems working with coded modulation/demodulation schemes. Prospective work includes the implementation and evaluation of such schemes in our prototype during real operation.
\bibliographystyle{IEEEtran}
\bibliography{references}
\end{document}